\newcommand{\ce}{\mathop{\mathrm{ce}} \nolimits}
\newcommand{\se}{\mathop{\mathrm{se}} \nolimits}

\documentclass[prl,superscriptaddress,twocolumn,showpacs]{revtex4}

\usepackage{amsfonts,amssymb,amsbsy,bm,subeqnarray,graphicx,amsmath}

\begin{document}

\title{Minimum uncertainty measurements of angle and
angular momentum}

\author{Z. Hradil}
\affiliation{Department of Optics,
Palacky University, 17. listopadu 50,
772 00 Olomouc, Czech Republic}

\author{J. \v{R}eh\'{a}\v{c}ek}
\affiliation{Department of Optics,
Palacky University, 17. listopadu
50, 772 00 Olomouc, Czech Republic}

\author{Z. Bouchal}
\affiliation{Department of Optics,
Palacky University, 17. listopadu
50, 772 00 Olomouc, Czech Republic}

\author{R. \v{C}elechovsk\'y}
\affiliation{Department of Optics,
Palacky University, 17. listopadu
50, 772 00 Olomouc, Czech Republic}

\author{L. L. S\'{a}nchez-Soto}
\affiliation{Departamento de \'Optica,
Facultad de F\'{\i}sica,
Universidad Complutense, 28040~Madrid, Spain}

\date{\today}

\begin{abstract}
The uncertainty relations for angle and angular
momentum are revisited. We use the exponential
of the angle instead of the angle itself and
adopt dispersion as a natural measure of
resolution. We find states that minimize the
uncertainty product under the constraint of a
given uncertainty in angle or in angular momentum.
These states are described in terms of  Mathieu
wave functions and may be approximated by a
von Mises distribution, which is the closest
analogous of the Gaussian on the unit circle.
We report experimental results using beam
optics that confirm our predictions.
\end{abstract}

\pacs{03.67.-a, 42.50.Dv, 42.50.Vk}

\maketitle

Light carries and transfers energy as well as linear
and angular momentum. The angular momentum contains
a spin contribution, associated with polarization,
and an orbital component, linked with the spatial
profile of the light intensity and phase~\cite{ABP2003}.

The seminal paper of Allen \textit{et al}~\cite{ABSW1992}
firmly establishes that the Laguerre-Gauss modes,
typical of cylindrical symmetry, carry a well-defined
angular momentum per photon. In the paraxial limit,
this orbital component is polarization independent
and arises solely from the azimuthal phase dependence
$e^{i m \phi}$, which gives rise to spiral wave fronts.
The index $m$ takes only integer values and can be seen
as the eigenvalue of the orbital angular momentum
operator. In consequence, the Laguerre-Gauss modes
constitute a complete set and can be used to represent
quantum photon states~\cite{EN1992,Wun2004,CPB2005}.

The possibility of exploiting these light fields for
driving micromachines, and their applications as
optical tweezers and traps, have attracted a good
deal of attention~\cite{SDAP1997,GO2001,PMA2001}.
Moreover, entangled photons prepared in a superposition
of states bearing a well-defined orbital angular
momentum provide access to multidimensional
entanglement. This is of considerable importance
in quantum information and cryptography because
with these states more information can be stored
and there is less sensitivity to
decoherence~\cite{MVWZ2001,MTT2002,VPJ2003,MVRH2004,AOEW2005}.

Recently, by precise measurements on a light
beam, an experimental test of the uncertainty
principle for angle and angular momentum has been
demonstrated~\cite{FBY2004,PBZ2005}. The idea is
to pass the beam through an angular aperture and
measure the resulting distribution of angular-momentum
states~\cite{LPB2002}. Moreover, one can even
identify the form of the aperture that corresponds
to the minimum uncertainty states for these variables.

In the following, we deal with cylindrical symmetry;
we are concerned with the planar rotations
by an angle $\phi$ in the plane $x$-$y$, generated
by the angular momentum along the $z$ axis, which for
simplicity will be denoted henceforth as $\hat{L}$.
In this respect, we recall that the proper definition
of angular variables in quantum mechanics is beset by
well-known difficulties~\cite{PLP1998,LS2000}. For the
case of a harmonic oscillator, the problems essentially
arise from two basic sources: the periodicity and the
semiboundedness of the energy spectrum. The first
prevents the existence of a phase operator in the
infinite-dimensional Hilbert space, but not of its
exponential. The second entails that this exponential
is not unitary.

Although we have here the same kind of problems
linked with the periodicity, the angular momentum
has a spectrum that includes both positive and
negative integers, which allows us to introduce a
well behaved exponential of the angle operator,
denoted by $\hat{E}$~\cite{LS1998}. Since the angle
is canonically conjugate to $\hat{L}$, we start
from the commutation relation~\cite{LS2000}
\begin{equation}
\label{Ler}
[ \hat{E},  \hat{L}] = \hat{E} .
\end{equation}
The goal of this Letter is precisely to develop a
comprehensive approach to the minimum uncertainty
states associated with the relation~(\ref{Ler}).
Our results will corroborate that the use of $E$
provides a good description of the angular behavior
and the associated minimum uncertainty states turn
out to be Mathieu beams in wave optics. This will
establish a proper basis for information processing
with these conjugate variables.

We also stress that, since angle is $2\pi$ periodic, the
corresponding quantum statistical description should also preserve
this periodicity. Provided that a non-periodic measure of the
angular spread is used, as for example the variance, such a
resolution depends on the $2\pi$ window chosen. To prevent this,
we recall that another appropriate and meaningful measure of
angular spread is the dispersion~\cite{Rao1965}
\begin{equation}
\label{dispersion}
D^2 = 1 - | \langle e^{i \phi} \rangle |^2.
\end{equation}
Here
\begin{equation}
\langle e^{i \phi} \rangle = \int_{0}^{2 \pi}
d\phi \ P(\phi) \
e^{i\phi}
\end{equation}
and $P(\phi)$ is the angle distribution. As expected,
it  possesses all the good properties: it is periodic,
the shifted distributions $P (\phi + \phi^\prime)$
are characterized by the same resolution, and for
sharp angle distributions it coincides with the
standard variance since $| \langle e^{i \phi} \rangle |^2
\simeq 1 + \langle \phi^2 \rangle$. In consequence,
the statistics of the exponential of the angle
provides a sensible measure of the angle resolution.

The action of the unitary operator $\hat{E}$ in the angular
momentum basis is
\begin{equation}
\hat{E} | m \rangle = | m -1 \rangle ,
\end{equation}
where the integer $m$  runs from $- \infty$ to $+ \infty$.
Therefore, $\hat{E}$ possesses a simple optical
implementation by means of phase mask removing a
charge + 1 from a vortex beam. The normalized
eigenvectors of $\hat{E}$ are
\begin{equation}
|\phi \rangle = \frac{1}{\sqrt{2 \pi}}
\sum_{m=- \infty}^\infty e^{i m \phi} | m \rangle ,
\end{equation}
and, in the representation they generate, we can write
\begin{equation}
\label{Lz}
\hat{L} = -i \frac{d}{d \phi},
\qquad \qquad
\hat{E} = e^{i \phi},
\end{equation}
which formally verify the fundamental relation (\ref{Ler}).

Let us  turn to the corresponding uncertainty relation.
When the standard form $(\Delta \hat{A} )^2 (\Delta \hat{B} )^2
\ge | \langle [\hat{A}, \hat{B} ] \rangle |^2/4$ is applied
to Eq.~(\ref{Ler}) and the previous notion of dispersion is
used, we get
\begin{equation}
\label{disp}
D^2  \ (\Delta \hat{L} )^2 \ge \frac{1}{4}(1-D^2) .
\end{equation}
This can be recast in terms of the cosine and sine
operators, $\hat{C} = (\hat{E} + \hat{E}^\dagger )/2$
and $\hat{S} = (\hat{E} - \hat{E}^\dagger )/2i$,
yielding
\begin{equation}
\label{unCS}
( \Delta \hat{C} )^2 (\Delta \hat{L} )^2 \ge
\frac{1}{4} |\langle \hat{C} \rangle |^2 ,
\qquad
( \Delta \hat{S} )^2 ( \Delta \hat{L} )^2 \ge
\frac{1}{4} |\langle \hat{S} \rangle |^2 .
\end{equation}
States satisfying the equality in an uncertainty
relation are sometimes referred to as intelligent
states~\cite{BL1981}. However, in the case of
Eq.~(\ref{disp}), the inequality cannot be saturated
(except for some trivial cases), since this would imply
to saturate both relations in (\ref{unCS}) simultaneously.
In other words, the formulation (\ref{disp}) is true
but too weak.

To get a saturable lower bound we look instead at
normalized states that minimize the uncertainty
product $D^2 \ (\Delta \hat{L})^2$ either for a
given $D^2$ or for a given $(\Delta \hat{L})^2$.
These have been called constrained minimum
uncertainty-product states~\cite{FBY2004}.
We approach this problem by the method of undetermined
multipliers. The linear combination of variations
[whether we minimize $D^2 \ (\Delta \hat{L})^2$
for a fixed $D^2$ or for a fixed $(\Delta \hat{L})^2$]
lead to the basic equation
\begin{equation}
\label{eigcom}
[ \hat{L}^2 +  \mu \hat{L} + ( q^\ast \hat{E} + q
E^\dagger )/2 ] | \Psi \rangle = a | \Psi \rangle ,
\end{equation}
where $\mu$, $q$, and $a$ are Lagrange multipliers. We
shall solve this eigenvalue equation in the angle
representation $\Psi (\phi) = \langle \phi | \Psi \rangle$.
Note first that, without loss of generality, we can
restrict ourselves to states with $\langle \hat{L} \rangle
= 0$, since we readily obtain solutions with mean
angular momentum $\bar{m}$ by multiplying the wave
function by $\exp(i \bar{m} \phi)$. Alternatively, we
observe that the change of variables $\exp(i \mu \phi)
\Psi (\phi)$ eliminates the linear term from (\ref{eigcom}).
In addition, we can take $q$ to be a real number,
since this only introduces an unessential global
phase shift. We therefore look at solutions of
\begin{equation}
\frac{d^2 \Psi (\phi)}{d\phi^2} -
(a - q \cos \phi ) \  \Psi (\phi) = 0 .
\end{equation}
To properly interpret this eigenvalue problem we
introduce the rescaled angular variable $ \eta =
\phi/2$, so that
\begin{equation}
\label{Mathieu}
\frac{d^2 \Psi (\eta)}{d\eta^2}  + [ a - 2 q
\cos (2 \eta) ] \ \Psi (\eta) = 0 ,
\end{equation}
which is the standard form of the Mathieu
equation~\cite{McL1947}. The variable $\eta$ has
a domain $ 0 \le \eta < 2 \pi$ and plays the role
of polar angle in elliptic coordinates. In our case,
the required periodicity imposes that the only
acceptable Mathieu functions are those periodic
with period of $\pi$ or $2 \pi$. The values of $a$
in Eq.~(\ref{Mathieu}) that satisfy this condition
are the eigenvalues of this equation. We  have then
two families of independent solutions, namely the
even and the odd angular Mathieu functions:
$\ce_n ( \eta, q) $ and $\se_n (\eta, q)$ with
$n = 0, 1, 2, \ldots .$, which are usually known
as the elliptic cosine and sine, respectively.
For $\ce_n (\eta, q)$  the eigenvalues are
denoted as $a_n (q)$, whereas for $\se_n (\eta, q)$
they are represented as $b_n (q)$: they form an
infinite set of countable real values that have the
property $a_0 < b_1 < a_1 < b_2 < \ldots$. The parity
and periodicity of these functions are exactly the
same as their trigonometric counterparts; that is,
$\ce_n (\eta, q)$ is even and $\se_n (\eta, q)$ is
odd in $\eta$, and they have period $\pi$ when $n$
is even or period $2 \pi$ when $n$ is odd.

Since the $2\pi$ periodicity in $\phi$ requires $\pi$
periodicity in $\eta$, the acceptable solutions for our
eigenvalue problem are the independent Mathieu functions
of even order
\begin{equation}
\label{Matsol}
\Psi_{2n} (\eta, q) = \sqrt{\frac{2}{\pi}}
\left \{
\begin{array}{cc}
\ce_{2n} (\eta, q) , &  \\
& \qquad n = 0, 1, \ldots, \\
\se_{2n} (\eta, q) ,
\end{array}
\right .
\end{equation}
where the numerical factor ensures a proper normalization,
according to the properties of these functions. In what
follows we shall consider only even solutions $\ce_{2n}
(\eta, q)$, although a parallel treatment can be done
for the odd ones.  After some calculations, we get
\begin{eqnarray}
(\Delta \hat{L})_{2n}^2 & = &
\displaystyle
\frac{1}{2 \pi}  \int_{0}^{\pi} d\eta \
\left[ \frac{d}{d\eta} \ce_{2n} (\eta, q) \right ]^2 =
 \frac{1}{4} [ A_{2n}(q) - 2q \Theta_{2n} (q) ] ,
\nonumber \\
& & \\
D_{2n}^2 & = & 1 - \frac{2}{\pi}  \left |
\int_{0}^{\pi} d\eta \ \ce_{2n}^2 (\eta, q)  \
\cos (2\eta) \right |^2 =   1 - | \Theta_{2n} (q) |^2 ,
\nonumber
\end{eqnarray}
where we have expanded the periodic functions $\ce_{2n}
(\eta, q)$ in Fourier series
\begin{equation}
\label{Fser}
\ce_{2n} (\eta, q) = \sum_{k=0}^\infty A_{2k}^{(2n)} (q)
\cos (2 k \eta) ,
\end{equation}
and we have integrated term by term, in such a way that
\begin{equation}
\Theta_{2n} (q) =  A^{(2n)}_{0} (q) A^{(2n)}_{2} (q) +
\sum_{k=0}^\infty A^{(2n)}_{2k} (q) A^{(2n)}_{2k +2} (q) .
\end{equation}
The coefficients $A_{2k}^{(2n)}$  determine the Fourier
spectrum and satisfy recurrence relations that are easily
obtained by substituting (\ref{Fser}) in the Mathieu
equation and can be efficiently computed by a variety
of methods~\cite{FP2001}.

If we expand $\ce_{2n} (\eta, q)$ in powers of $q$
and retain only linear terms~\cite{McL1947}, we
have
\begin{eqnarray}
(\Delta \hat{L})_{2n}^2 & = &
\displaystyle
\frac{(2 n)^2}{4} +  \frac{4n^4 - 3n^2 +1}
{8 (4n^2 - 1)^2} q^2 , \nonumber \\
& & \\
D_{2n}^2 & = & 1 - \frac{1}{4 (4n^2 - 1)^2} q^2 ,
\nonumber
\end{eqnarray}
which shows a quadratic increasing with $q$ of the
angular-momentum variance and a decreasing of the
angle dispersion. The uncertainty product, up to terms
$q^2$, reads as
\begin{equation}
D_{2n}^2 \ (\Delta \hat{L})_{2n}^2  = n^2  +
\frac{1}{4} [ (4n^4 - 5n^2 +1) (1 - D_{2n}^2 ) ] .
\end{equation}
It is clear that this product attains its minimum
value for the fundamental mode $n=0$, which saturates
the bound in Eq.~(\ref{disp}) for this range of
values of $q$.

Note that for large dispersions ($q\rightarrow 0$)
the fundamental wave function may be approximated by
\begin{equation}
\label{mises}
P_0 (\phi ) \propto | \ce_0 (\eta, q) |^2
\simeq \exp(- q \cos \phi), \qquad  (q \rightarrow 0) ,
\end{equation}
which is the von Mises distribution, also known as the
normal distribution on the unit circle~\cite{Fis1995}.
This remarkable result shows that optimal states are
very close to Gaussians on the unit circle~\cite{Bre1985}.
Curiously enough, it has been recently found that the
von Mises distribution maximizes the entropy for a fixed
value of the dispersion~\cite{LR2000}.

\begin{figure}
\includegraphics[width=0.90\columnwidth]{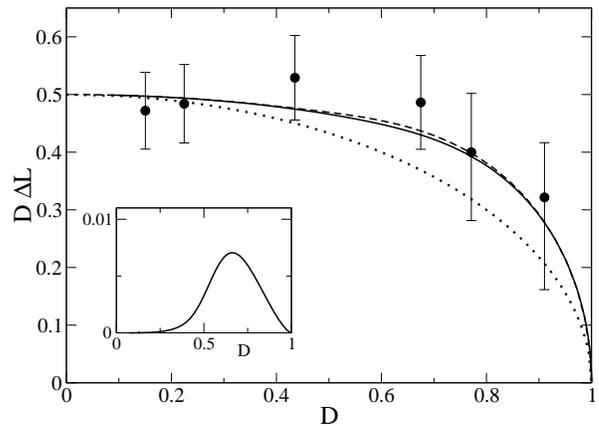}
\caption{\label{fig:exper}
Theoretical and experimentally observed uncertainty
products as a function of the dispersion. Solid
line represents the fundamental Mathieu beam
while the broken line represents the von Mises
approximation. The difference between these
two wave functions appears plotted in the inset.
We have included also the ideal bound given
by Eq.~(\ref{disp}).}
\end{figure}

In the opposite limit of small dispersions ($q \rightarrow
\infty$), one can also check that
\begin{equation}
P_0 (\phi ) \propto | \ce_0 (\eta, q) |^2
\simeq \exp(- \sqrt{q} \cos \phi), \qquad (q \rightarrow \infty).
\end{equation}
Therefore, von Mises wave functions constitute an
excellent approximation to the minimum-uncertainty
Mathieu wave functions, except perhaps for intermediate
values of the dispersions, where a deviation may occur.
In Fig.~\ref{fig:exper} we have plotted  $D \
(\Delta \hat{L} )$ in terms of $D$: the solid line
represents the fundamental Mathieu beam, which provides
the optimal angular resolution, while the broken line
represents the von Mises approximation.  The very small
difference between these two curves is magnified in the
inset. For the purposes of comparison, the ideal bound
coming from Eq.~(\ref{disp}) is plotted as a dotted line.
We stres that the minimum uncertainty states with large
dispersions present wide angular distributions and
{\em vice versa}.

\begin{figure}
\includegraphics[width=0.95\columnwidth]{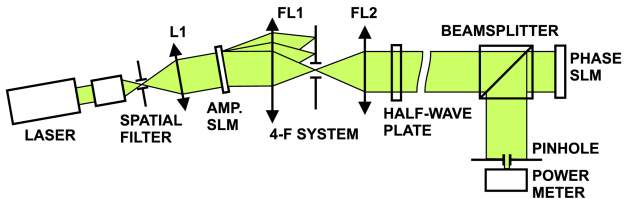}
\caption{\label{fig:setup}
Experimental setup for the generation of
beams with a von Mises distribution and subsequent
detection of the associated angular-momentum components.}
\end{figure}

This theoretical approach can be experimentally realized,
although our capabilities to prepare states and perform
measurements on demand are limited by the present technology.
Figure~\ref{fig:setup} shows our experimental setup. Two
spatial light modulators (SLM) were used: the amplitude
SLM (CRL Opto, $1024 \times 768$ pixels) generates the
angular-restricted light beams, while the phase SLM
(Boulder, $512 \times 512$ pixels) works as an analyzing
hologram.  The beam generated by an Ar laser (514 nm, 200 mW)
is spatially filtered, expanded and collimated by the
lens L$_1$ and impinges on the hologram generated by
the amplitude SLM. The bitmap of  the hologram is
computed as an interference pattern of the signal beam
$U_s$ and an inclined  reference plane wave $U_p =
u_p \exp[-ik(x \sin \gamma + z \cos \gamma)]$, where
$k$ denotes the wave number and $\gamma$ is the
angle of the transversal component of $\mathbf{k}$
with respect to the $z$ axis. After illuminating the
hologram with the collimated beam, which can be
approximated by the plane wave $U_i =  u_i \exp(-ikz)$,
the field behind the SLM can be written as $U_t =
U_i |U_s + U_p|^2$. The Fourier spectrum of this
transmitted beam is localized at the back focal plane
of the first Fourier lens FL$_1$ and consists of three
diffraction orders $(-1, 0, +1)$. The undesired 0 and $-1$
orders are removed by the spatial filter. After inverse
Fourier transformation, performed by the second Fourier
lens FL$_2$, a collimated beam with the required complex
amplitude profile $U_s$ is obtained. This field impinges
on the reflecting phase SLM. The hologram on the SLM is
of the form $U_h=\exp[i m \phi + iS(x)]$, where $\phi$ is
the azimuthal angle and $S(x)=\mod(x, \Lambda_x)$
is the sawtooth function of period $\Lambda_x$, which
ensures deflection of the undesired orders coming
from the pixel structure of the SLM.

When the field impinging on the phase SLM has a structure
$\sum_{n=-\infty}^{\infty} a_n \exp(i n \phi)$, the
hologram modifies its helicity so that the reflected beam
can be written as $\sum_{n=-\infty}^{\infty} a_n
\exp[ i (n+m) \phi]$. The components for which $n+m=0$
present no helicity and their intensity profile consists
of a light spot localized at the center of the beam. The
other components have a nonzero helicity and the center
of the intensity pattern remains dark. From this fact,
the weight coefficients of the superposition can be
determined by selective intensity measurements performed
by a pinhole and a power meter.

A Laguerre-Gauss beam was used to align the setup and
subsequently a beam with a von Mises distribution (\ref{mises})
of the transversal amplitude was generated. Each angular
amplitude width was scanned for values of the helicities
from $n=-20$ to $n=20$. Experimentally measured
uncertainty products are depicted in Fig.~\ref{fig:exper}
by solid circles.

Given the accuracy of the measurements (indicated by
error bars in Fig.~\ref{fig:exper}), they fit quite
well the theoretical predictions. Our present experiment
distinguishes between the uncertainty product of optimal
states and the ideal limit. It is, however, not possible
to dicriminate between the Mathieu and von Mises beams.
Keeping in mind that von Mises and Mathieu states play the
same role for the spatial degrees of freedom as Gaussian
states for quadratures, the observation of the nonclassical
behavior of angle and angular momentum is a challenging
problem left for future studies.

In conclusion, we have formulated the uncertainty relations
for angle and angular momentum based on dispersion as a
correct statistical measure of error. The optimal states
were derived and identified with the Mathieu wave functions.
An optical test of the derived uncertainty relations was
proposed and performed experimentally.

We acknowledge discussions with Hubert de Guise. This work
was supported by the Czech Ministry of Education, Project
MSM6198959213, the Czech Grant Agency, Grant  202/06/307,
and the Spanish Research Directorate, Grant FIS2005-06714.

\end{document}